\documentclass[twocolumn]{svjour3}          
\smartqed  
\usepackage{graphicx}
\usepackage{natbib}
\usepackage{amsmath,color}
 \journalname{Computational Economics}
\begin{document}

\title{Computational experiments successfully predict the emergence of autocorrelations in ultra-high-frequency stock returns}

\subtitle{Published: Computational Economics 50 (4), 579-594 (2017)}

\titlerunning{Computational experiments predict emergence of autocorrelations}        

\author{Jian Zhou  \and
        Gao-Feng Gu  \and
        Zhi-Qiang Jiang  \and
        Xiong Xiong  \and
        Wei Chen  \and
        Wei Zhang  \and
        Wei-Xing Zhou
}


\institute{Jian Zhou \at
              School of Business, East China University of Science and Technology, Shanghai 200237, China
           \and
           Gao-Feng Gu, Zhi-Qiang Jiang \at
              School of Business and Research Center for Econophysics, East China University of Science and Technology, Shanghai 200237, China
           \and
           Wei Chen \at
              Shenzhen Stock Exchange, 5045 Shennan East Road, Shenzhen 518010, China
           \and
           Xiong Xiong, Wei Zhang \at
              College of Management and Economics and China Center for Social Computing and Analytics, Tianjin University, Tianjin 300072, China \\
              \email{xxpeter@tju.edu.cn}           
           \and
           Wei-Xing Zhou \at
              School of Business, Department of Mathematics, and Research Center for Econophysics, East China University of Science and Technology, Shanghai 200237, China \\
              \email{wxzhou@ecust.edu.cn}           
}

\date{Received: 09 September 2015 / Accepted: 11 August 2016 / Published online: 24 August 2016}

\maketitle

\begin{abstract}
  Social and economic systems are complex adaptive systems, in which heterogenous agents interact and evolve in a self-organized manner, and macroscopic laws emerge from microscopic properties. To understand the behaviors of complex systems, computational experiments based on physical and mathematical models provide a useful tools. Here, we perform computational experiments using a phenomenological order-driven model called the modified Mike-Farmer (MMF) to predict the impacts of order flows on the autocorrelations in ultra-high-frequency returns, quantified by Hurst index $H_r$. Three possible determinants embedded in the MMF model are investigated, including the Hurst index $H_s$ of order directions, the Hurst index $H_x$ and the power-law tail index $\alpha_x$ of the relative prices of placed orders. The computational experiments predict that $H_r$ is negatively correlated with $\alpha_x$ and $H_x$ and positively correlated with $H_s$. In addition, the values of $\alpha_x$ and $H_x$ have negligible impacts on $H_r$, whereas $H_s$ exhibits a dominating impact on $H_r$. The predictions of the MMF model on the dependence of $H_r$ upon $H_s$ and $H_x$ are verified by the empirical results obtained from the order flow data of 43 Chinese stocks.
\end{abstract}
\vspace{12pt}

\section{Introduction}
\label{S1:Introduction}

Social and economic systems consist of interacting heterogenous agents. Macroscopic laws and collective behaviors emerge from the self-organized evolution of these complex systems. Theories in the top-down framework frequently fail in the description of complex socioeconomic phenomena and have very limited predictive power \citep{Farmer-Foley-2009-Nature}. For instance, the so-called economic men ({\textit{Homo Oeconomicus}}) in economics are hypothesized to be homogenous, do not interact or interact in a linear way, and pursue to maximizing their profits rationally, like ideal gas in physics. However, agents in real societies are heterogenous and interact in a nonlinear manner \citep{Schweitzer-Fagiolo-Sornette-VegaRedondo-Vespignani-White-2009-Science,Lux-2011-Nature}. A bottom-up phenomenological framework in natural sciences should be adopted in the study of social sciences \citep{Bouchaud-2008-Nature,Lux-Westerhoff-2009-NP}, which is now known as computational social science \citep{Lazer-Pentland-Adamic-Aral-Barabasi-etc-2009-Science}.

Indeed, computational experiments are important and well-recognized in the fields of econophysics and sociophysics \citep{Farmer-Foley-2009-Nature,Sornette-2014-RPP}. A general framework is as follows. First, one performs phenomenological analysis to uncover important statistical laws at the microscopic level. Second, one constructs a model based on the microscopic laws. Third, one performs numerical simulations to generate macroscopic properties of the complex system. Forth, if the simulation results deviate from the stylized facts of the real system, one needs to go to the first step to find out possible missing ingredients and then improve the model \citep{Gu-Zhou-2009-EPL,Li-Zhang-Zhang-Zhang-Xiong-2014-IS}. Once the model is in store, one can perform computational experiments from the scenario-response perspective. In other words, computational experiments have predictive powers by projecting the responses of a system to external stimuli under certain scenarios and can thus provide guidance to policy makers \citep{Farmer-Foley-2009-Nature}. Alternatively, computational experiments are able to unravel uncovered properties of a system, which are testable using empirical analysis.

In this work, we perform computational experiments using an empirical order-driven model to identify the microscopic determinants of the autocorrelation structure of stock return time series. Stock markets are efficient in the weak form if no significant autocorrelations can be identified in the returns. Overwhelming evidence shows that the weak-form efficiency holds for intraday high-frequency returns and low-frequency (daily, weekly, monthly, et al.) returns in the sense that the Hurst index of returns ($H_r$) is insignificantly different from 0.5. However, it is not clear if $H_r$ differs from 0.5 or not at the transaction level. In addition, the microscopic mechanisms leading to specific $H_r$ values are unclear. We aim at understanding the impacts of order flows on the weak-form efficiency through computational experiments based on an empirical order-driven model -- the modified Mike-Farmer (MMF) model. The seminal model was proposed by \cite{Mike-Farmer-2008-JEDC}, mimicking the processes of order placement and cancellation. The Mike-Farmer (MF) model is able to reproduce the main stylized facts of stock returns, such as power-law tails in the return distribution and absence of long memory in return time series. However, the MF model fails to reproduce the phenomenon of volatility clustering \citep{Mike-Farmer-2008-JEDC}. \cite{Gu-Zhou-2009-EPL} proposed a modified MF (MMF) model, which successfully reproduces all these stylized facts. There are three possible determinants embedded in the MMF model, including the tail heaviness of relative prices of the placed orders characterized by the tail index $\alpha_x$, the degree of long memory in relative prices quantified by its Hurst index $H_x$, and the strength of long memory in order directions depicted by $H_s$. We investigate the impacts of these variables on the correlation structure of return time series.

Our work is directly related to the Efficient Markets Hypothesis (EMH), which is one of the cornerstones of modern finance \citep{Fama-1970-JF,Fama-1991-JF}. There are three major versions of the hypothesis: weak form, semi-strong form, and strong form \citep{Fama-1970-JF,Fama-1991-JF}. The weak form efficiency hypothesis suggests that asset prices are unpredictable using historical prices. The study of the weak-form EMH can be traced back to \cite{Bachelier-1900}, who suggests that speculative prices follow random walks. Early empirical evidence and theoretical analysis support the random walk hypothesis \citep{Cowles-1933-Em,Working-1934-JASA,Kendall-1953-JRSSA,Osborne-1959a-OR,Cootner-1964,Samuelson-1965-IMR,Mandelbrot-1966-JB}. A classical approach to test the random walk hypothesis is to calculate the Hurst index $H$ of return time series \citep{Mandelbrot-1971-Em}. A time series is uncorrelated if its Hurst index $H=0.5$, antipersistent if $H<0.5$, or persistent if $H>0.5$ \citep{Mandelbrot-VanNess-1968-SIAMR}. Different methods have been utilized to estimate Hurst indexes of financial return time series and controversial results are reported \citep{Cajueiro-Tabak-2004a-PA,AlvarezRamirez-Alvarez-Rodriguez-FernandezAnaya-2008-PA,Cajueiro-Tabak-2008-CSF,Mishra-Sehgal-Bhanumurthy-2011-RFE,Jiang-Xie-Zhou-2014-PA}. We note that a well-designed statistical method is necessary to draw a conclusion on the presence of long memory in returns. It is more likely that returns are uncorrelated in long term and show inefficiency in certain short time periods \citep{Jiang-Xie-Zhou-2014-PA}. However, we are not going to devote to this debate. Rather, we attempt to identify possible microscopic determinants in order flows that impact the correlation structure of financial return time series.

Our work is also related to the microstructure literature. \cite{Cont-Kukanov-Stoikov-2014-JFEm} find that, over short time intervals, price changes are mainly driven by the order flow imbalance, defined as the imbalance between supply and demand at the best bid and ask prices. It has also been shown that market participants who function at this very high frequency level look for this autocorrelation in order flows to design trading strategies \citep{Clark-Joseph-2013-WP,Fishe-Haynes-Onur-2015-SSRN}, which is actually based on herding mechanisms. Although market participants cannot make high-frequency trading due to the $T+1$ rule, our research still provides some insights on the dynamics of Chinese stocks.

\section{Description of methods and data sets}

\subsection{Model description}
The order-driven model proposed by \cite{Mike-Farmer-2008-JEDC} mimics the processes of order placement and order cancellation. An order is determined by its direction, price and size.
\begin{enumerate}
  \item[(1)] {\textbf{Order direction.}} We denote the order direction by ``$+1$'' if it is a buy or ``$-1$'' if it is a sell. The time series of order direction has long memory characterized by a Hurst index $H_s$ \citep{Lillo-Farmer-2004-SNDE}. We can simulate the sequence of order direction using fractional Brownian motions with Hurst index $H_s$.
  \item[(2)] {\textbf{Order price.}} The logarithmic order price $\pi(t)$ at event time $t$ is determined by the relative price $x(t)$. One defines $x(t)=\pi(t)-\pi_b(t-1)$ for buy orders and $x(t)=\pi_a(t-1)-\pi(t)$ for sell orders, where $\pi_b(t)=\ln{b(t)}$ and $\pi_a(t)=\ln{a(t)}$ with $a(t)$ and $b(t)$ being the best ask and bid prices right before the event time $t$. In the MF model, $x(t)$ follows a student distribution with the freedom degree being $\alpha_x$. In the MMF model, \cite{Gu-Zhou-2009-EPL} used an additional ingredient that the relative prices are long-term correlated with a Hurst index $H_x$.
  \item[(3)] {\textbf{Order size.}} For simplicity, all orders are set to have the same size \citep{Mike-Farmer-2008-JEDC,Gu-Zhou-2009-EPL,Meng-Ren-Gu-Xiong-Zhang-Zhou-Zhang-2012-EPL}.
\end{enumerate}

At each simulation step, we check if the unexecuted orders on the book are cancelled or not. Following \cite{Mike-Farmer-2008-JEDC}, for each order on the limit order book, its conditional cancellation probability $P=A(1-e^{-Y_i})(imb+B)/n_{\mathrm{tot}}$ is calculated, where $A=1.12$, $B=0.2$, $Y_i$ is the ratio of the current distance to execution to the initial distance to execution, $imb$ is the imbalance of the order book, and $n_{\mathrm{tot}}$ is the total number of orders on the order book. In this way, the processes of order placement and order cancellation can be simulated. Note that an excellent review of the statistical properties of limit order books is given in \cite{Gould-Porter-Williams-McDonald-Fenn-Howison-2013-QF}.

\subsection{Simulation setting}
We simulate the MMF model with different combinations of the three parameters, in which the Hurst indexes $H_s$ and $H_x$ range from 0.50 to 0.95 with a step of 0.05 and the tail exponent $\alpha_x$ varies from 1.00 to 1.65 also with a step of 0.05. For each combination $(H_s,H_x,\alpha_x)$, we simulate the MMF model for 100 repeated runs. In each run, we simulate $2 \times 10^5$ steps and record the return time series with a length of near $4 \times 10^4$ after removing the transient period, where the return is calculated as follows \citep{Mike-Farmer-2008-JEDC}:
\begin{equation}
   r(t) = \frac{\pi_a(t)+\pi_b(t)}{2} - \frac{\pi_a(t-1)+ \pi_b(t-1)}{2},
\end{equation}
which is the trade-by-trade price fluctuation after each transaction at time $t$ \citep{Lillo-Farmer-Mantegna-2003-Nature,Lim-Coggins-2005-QF,Zhou-2012-NJP,Zhou-2012-QF}. In our computational experiments presented in this work, the tick size is set to be 0.01.

To determine the Hurst index of each return time series, we adopted the detrended fluctuation analysis (DFA) proposed by \cite{Peng-Buldyrev-Havlin-Simons-Stanley-Goldberger-1994-PRE} (see also \cite{Kantelhardt-KoscielnyBunde-Rego-Havlin-Bunde-2001-PA}), which is one of the most efficient methods for the estimation of Hurst index \citep{Shao-Gu-Jiang-Zhou-Sornette-2012-SR}. However, the scaling behavior is not good for some time series. We thus turned to adopt the centred detrending moving average (DMA) method \citep{Alessio-Carbone-Castelli-Frappietro-2002-EPJB,Carbone-Castelli-Stanley-2004-PA,Carbone-Castelli-Stanley-2004-PRE,Xu-Ivanov-Hu-Chen-Carbone-Stanley-2005-PRE,Carbone-2009-IEEE,Gu-Zhou-2010-PRE,Jiang-Zhou-2011-PRE}, which gives better scaling behaviors for those cases. The better performance of the centred DMA does not contradict with the numerical results in \cite{Shao-Gu-Jiang-Zhou-Sornette-2012-SR} because the conclusions in \cite{Shao-Gu-Jiang-Zhou-Sornette-2012-SR} are based on fractional Gaussian noises. We partition each return time series into segments of length $\ell$ and determine the average DMA fluctuation $F(\ell)$ over all segments. The Hurst index is obtained from $F(\ell) \sim \ell^H$.

\begin{figure}
  \centering
  \includegraphics[width=0.95\textwidth,height=0.40\textwidth]{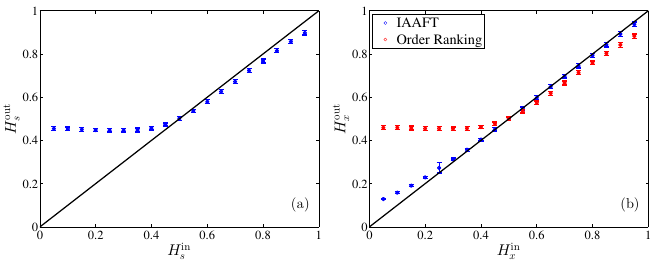}
  \caption{Comparison of the input and output Hurst indexes for the order sign series and the relative price series. (a) Input and output Hurst indexes for order sign series by taking the signs of of the increments of fractional Brownian motions. (b) Input and output Hurst indexes for the relative prices generated by the order ranking method and the iterative amplitude adjusted Fourier transform (IAAFT) approach.}
  \label{Fig:Hin:Hout}
\end{figure}

To generate orders, we need to synthesize time series of order signs and relative prices with given Hurst indices $H_s$ and $H_x$. For the series of order signs, we generate a fractional Brownian motion with Hurst index $H_s$ and obtain the sign sequence of its increments. A comparison of the output Hurst index $H_x^{\rm{out}}$ and the input Hurst index $H_x^{\rm{in}}$ in Fig.~\ref{Fig:Hin:Hout} (a) shows that the synthetic order signs are not reliable when $H_s$ is less than 0.5. For the series of relative prices, we compared the outputs of two widely used methods, the rank ordering method \citep{Bogachev-Eichner-Bunde-2007-PRL,Zhou-2008-PRE} and iterative amplitude adjusted Fourier transform (IAAFT) approach \citep{Schreiber-Schmitz-1996-PRL}. Fig.~\ref{Fig:Hin:Hout} (b) shows that the IAAFT approach provides much better results, but with deviations for small $H_x$.

\subsection{Data description}
We use the order flow data of 32 A-share stocks and 11 B-share stocks traded on the Shenzhen Stock Exchange (SZSE) in 2003. The A-shares are common stocks issued by mainland Chinese companies, subscribed and traded in the Chinese currency Renminbi, listed on mainland Chinese stock exchanges, bought and sold by Chinese nationals and approved foreign investors. The A-share market was open only to domestic investors in 2003. Note that the Chinese stock market is the largest emerging market in the world and became the second largest stock market after the USA market in 2009. Our sample stocks were part of the 40 constituent stocks included in the Shenzhen Stock Exchange component index in 2003. However, we were not able to retrieve the data for all 40 A-share constituents. The codes of stocks have six digital numbers. The codes are initiated with ``0'' for A-share stocks and ``2'' for B-share stocks. A company can be listed in both the A-share market and the B-share market. In this case, the only difference between the two codes is the first digit.

\section{Computational experiments}
\begin{figure}
  \centering
  \includegraphics[width=0.49\textwidth]{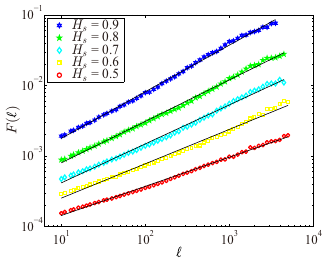}
  \caption{Dedrended fluctuation analysis of five typical return time series generated by the MMF model. In these cases, $\alpha=1.30$ and $H_x=0.8$ are fixed, while $H_s$ takes different values. The solid lines obtained from linear least-squares regressions show nice power-law relationships between $F(\ell)$ and $\ell$.}
  \label{Fig:Hx:l:Fl}
\end{figure}

Figure~\ref{Fig:Hx:l:Fl} shows the average fluctuation function $F(\ell)$ with respect to the segment size $\ell$ for five typical return time series. It is evident that these curves exhibit nice power-law relationships in the sense that the average fluctuation functions fluctuate closely around the corresponding fitted straight lines. We perform linear least-squares regressions of $\ln{F(\ell)}$ against $\ln\ell$ within the scaling range $10\leq\ell<4500$. The resulting slopes are the estimates of the Hurst indexes of the return time series. It is found that the slope $H_r$ increases with $H_x$. For each combination of model parameters, the 100 Hurst indexes of the 100 return time series in repeated simulations are computed. We use all these Hurst indexes, denoted by $H_r(\alpha_x,H_x,H_s)$, for further analysis.

We calculate the three Pearson correlation coefficients of the dependent variable $H_r$ and the independent variables $\alpha_x$, $H_s$ and $H_x$. In calculating the correlation coefficient between $H_r$ and an independent variable, we ignore the influence of other two variables. We find that $\rho(H_r,\alpha_x)=-0.049$, $\rho(H_r,H_x)=-0.156$, and $\rho(H_r,H_s)=0.953$. All these correlation coefficients are significantly different from zero at the significance level of 0.001. It suggests that the Hurst index $H_r$ of the returns is negatively correlated with $\alpha_x$ and $H_x$ and positively correlated with $H_s$. In addition, the value of $H_s$ plays the most influencing role on $H_r$, consistent with the results in \cite{Gu-Zhou-2009-EPL}.

\setlength\tabcolsep{2.0pt}
\begin{table*}
  \centering
  \caption{\label{TB:Hr:Hx:Hs:alpha1.3} Estimated Hurst index $H_r(\alpha_x,H_x,H_s)$ of return time series generated from the MMF model for fixed $\alpha=1.3$. The numbers in parentheses are the standard deviations multiplied by 100.}
  \medskip
\begin{tabular}{cc|*{10}{c}}\hline\hline
& & \multicolumn{10}{c}{$H_x$}\\\cline{3-12}
& &{0.50} &{0.55} &{0.60}  &{0.65} &{0.70} &{0.75} &{0.80} &{0.85} &{0.90} &{0.95}\\\hline
        & 0.50 & $0.46(1)$ & $0.45(1)$ & $0.45(1)$ & $0.45(1)$ & $0.44(1)$ & $0.44(1)$ & $0.43(1)$ & $0.43(2)$ & $0.42(2)$ & $0.40(2)$\\
        & 0.55 & $0.47(1)$ & $0.47(1)$ & $0.46(1)$ & $0.46(1)$ & $0.45(1)$ & $0.45(1)$ & $0.44(1)$ & $0.44(2)$ & $0.43(2)$ & $0.41(2)$\\
        & 0.60 & $0.49(1)$ & $0.49(1)$ & $0.48(1)$ & $0.48(1)$ & $0.47(1)$ & $0.47(1)$ & $0.46(1)$ & $0.46(1)$ & $0.45(2)$ & $0.43(2)$\\
        & 0.65 & $0.52(1)$ & $0.52(1)$ & $0.51(1)$ & $0.51(1)$ & $0.50(1)$ & $0.50(1)$ & $0.49(1)$ & $0.48(1)$ & $0.47(2)$ & $0.46(2)$\\
  $H_s$ & 0.70 & $0.55(1)$ & $0.55(1)$ & $0.55(1)$ & $0.54(1)$ & $0.54(1)$ & $0.53(2)$ & $0.52(1)$ & $0.52(1)$ & $0.51(2)$ & $0.49(2)$\\
        & 0.75 & $0.58(1)$ & $0.58(1)$ & $0.58(1)$ & $0.58(2)$ & $0.57(1)$ & $0.57(1)$ & $0.56(1)$ & $0.55(1)$ & $0.54(2)$ & $0.53(2)$\\
        & 0.80 & $0.61(1)$ & $0.61(1)$ & $0.61(1)$ & $0.61(1)$ & $0.60(1)$ & $0.60(1)$ & $0.60(1)$ & $0.59(1)$ & $0.58(2)$ & $0.57(2)$\\
        & 0.85 & $0.63(1)$ & $0.63(1)$ & $0.63(1)$ & $0.63(1)$ & $0.63(1)$ & $0.63(1)$ & $0.62(1)$ & $0.62(1)$ & $0.61(2)$ & $0.60(2)$\\
        & 0.90 & $0.64(2)$ & $0.64(1)$ & $0.64(1)$ & $0.64(1)$ & $0.64(1)$ & $0.64(1)$ & $0.64(1)$ & $0.64(2)$ & $0.64(2)$ & $0.62(2)$\\
        & 0.95 & $0.64(1)$ & $0.63(1)$ & $0.64(1)$ & $0.64(2)$ & $0.64(2)$ & $0.65(1)$ & $0.65(1)$ & $0.64(1)$ & $0.65(1)$ & $0.64(2)$\\\hline\hline
\end{tabular}
\end{table*}

\begin{figure*}
  \centering
  \includegraphics[width=0.95\textwidth]{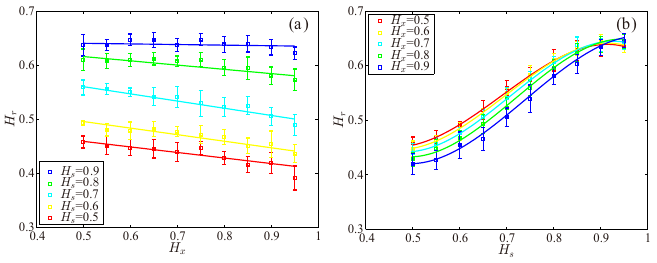}
  \caption{Dependence of $H_r$ on $H_x$ for different $H_s$ values (a) and on $H_s$ for different $H_x$ values (b) with fixed $\alpha_x=1.3$. The lines in plot (a) are the linear fits to the data points, while the solid curves in plot (b) are fits with three-order polynomials.}\label{Fig:Hr:Hx:Hs}
\end{figure*}

Because $\alpha_x$ seemingly has the weakest influence on $H_r$, we first study the impacts of $H_x$ and $H_s$ by fixing $\alpha_x=1.3$. The average Hurst indexes of the return time series are presented in Table~\ref{TB:Hr:Hx:Hs:alpha1.3}. The small values of the standard variations in parentheses show the robustness of the MMF model in generating return series. To have a better visibility, we plot the dependence of $H_r$ on $H_x$ for fixed $H_s$ in Fig.~\ref{Fig:Hr:Hx:Hs}(a) and on $H_s$ for fixed $H_x$ in Fig.~\ref{Fig:Hr:Hx:Hs}(b). We find that $H_r$ decreases with increasing $H_x$ and increases with increasing $H_s$. According to Fig.~\ref{Fig:Hr:Hx:Hs}(a), $H_r$ exhibits nice linear dependence with respect to $H_x$. The dependence between $H_r$ and $H_s$ in Fig.~\ref{Fig:Hr:Hx:Hs}(b) can also be roughly treated as being linear. We thus propose a linear dependence equation and obtain its coefficients as follows
\begin{equation}
  H_r = 0.23 - 0.08H_x + 0.52H_s,
  \label{Eq:Hr:Hx:Hs:Linear}
\end{equation}
where all the coefficients are significantly different from 0 and the adjusted $R^2$ is 0.935. A closer scrutiny of Fig.~\ref{Fig:Hr:Hx:Hs}(b) shows that the curves are nonlinear when $H_s$ is close to 0.5 and 0.95. We find that three-order polynomials can fit the data better than a linear function as depicted in Fig.~\ref{Fig:Hr:Hx:Hs}(b). This leads to the following equation:
\begin{equation}
    H_r =1.75 - 0.08H_x -6.23H_s +9.72H_s^2-4.55H_s^3,
  \label{Eq:Hr:Hx:Hs:Nonlinear}
\end{equation}
where all the coefficients are significantly different from 0 and the adjusted $R^2$ is 0.953.

When the tail exponent $\alpha_x$ of the relative prices is taken into consideration, we propose the following three-variable linear relationship
\begin{equation}
  H_r = 0.25 - 0.02 \alpha_x - 0.08H_x + 0.52H_s,
\label{Eq:Hr:alpha:Hx:Hs}
\end{equation}
where all the coefficients are significantly different from 0 and the adjusted $R^2$ is 0.935. It is interesting to notice that this equation is very similar to Eq.~(\ref{Eq:Hr:Hx:Hs:Linear}) and the coefficients for $H_x$ and $H_s$ are exactly the same in both equations. The only difference is that the intercept 0.23 in Eq.~(\ref{Eq:Hr:Hx:Hs:Linear}) is replaced by $0.25-0.02\alpha_x$ in Eq.~(\ref{Eq:Hr:alpha:Hx:Hs}). The positive and negative dependence between $H_r$ and the independent variables shown in Eq.~(\ref{Eq:Hr:alpha:Hx:Hs}) is consistent with the results from the Pearson correlation coefficients. When $\alpha_x=1.3$, $H_x=0.80$ and $H_s=0.75$, which are typical values for real stocks \citep{Mike-Farmer-2008-JEDC,Gu-Zhou-2009-EPL}, we have $H_r=0.55$, indicating that there is no long-term memory in the return time series.

We now investigate the robustness of the $H_r$ value with respect to the disturbance in $\alpha_x$, $H_x$ and $H_s$. According to Eq.~(\ref{Eq:Hr:alpha:Hx:Hs}), we have
\begin{equation}
  \Delta{H_r}=-0.02\Delta\alpha_x,
\end{equation}
when $H_x$ and $H_s$ are fixed. It means that, if $\alpha_x$ increases by 1, $H_r$ will decrease only by 0.02. Because the variation range of $\alpha_x$ is from 1.0 to 1.65 for stocks traded on the London Stock Exchange \citep{Mike-Farmer-2008-JEDC}, the variation of $H_r$ caused by $\alpha_x$ is less than 0.01. Hence, we can conclude that the tail heaviness characterized by $\alpha_x$ has a negligible impact on the Hurst index of the output return time series.

When $\alpha_x$ and $H_s$ are fixed, we have
\begin{equation}
  \Delta{H_r}=-0.08\Delta{H_x}.
\end{equation}
We note that $H_x$ is close to 0.8 for different stocks in developed market \citep{Zovko-Farmer-2002-QF} and emerging market \citep{Gu-Zhou-2009-EPL}. A value of $\Delta{H_x}=0.25$ will result in $\Delta{H_r}=-0.02$. This explains why both the MF model and the MMF model reach the same conclusion that the generated returns are not long-term correlated \citep{Mike-Farmer-2008-JEDC,Gu-Zhou-2009-EPL}. It means that the strength of autocorrelations in the relative price characterized by $H_x$ also has a minor impact on the Hurst index of the return time series.

When $\alpha_x$ and $H_x$ are fixed, we have
\begin{equation}
  \Delta{H_r}=0.52\Delta{H_s}.
\end{equation}
It is clear that the strength of long-term correlations $H_s$ in order direction has a dominating impact on $H_r$. For instance, if $H_s$ increases (or decreases) by 0.2, $H_r$ will increase (or decrease) by 0.1. Consider the real case that $\alpha_x=1.3$ and $H_x=0.8$. If there is no memory in order direction such that $H_s$ is close to 0.5, we have $H_r=0.42$ and the return time series is thus antipersistent. If the correlation in order direction is extremely strong such that $H_s\to1$, we have $H_r=0.68$ showing that the returns are persistent.

If we take the nonlinear behavior of $H_r$ against $H_s$ into consideration,  we can obtain the following equation:
\begin{equation}
  H_r=1.75-0.02\alpha_x-0.08H_x-6.11H_s+9.53H_s^{2}-4.45H_s^{3},
\label{eq:Hr:alpha:Hss}
\end{equation}
where all the coefficients differ significantly from 0 and the adjusted $R^2$ is 0.952. This equation fits the data with a slightly smaller error than Eq.~(\ref{Eq:Hr:alpha:Hx:Hs}). We find that the coefficients for $\alpha_x$ and $H_x$ are the same as in Eq.~(\ref{Eq:Hr:alpha:Hx:Hs}). If $\alpha_x=1.3$, $H_x=0.8$ and $H_s$ varies from 0.5 to 1.0, $H_r$ ranges from 0.43 to 0.63.

\section{Empirical validation}

We perform empirical analysis on each trading day using the order flow data of 32 A-share stocks and 11 B-share stocks traded on the Shenzhen Stock Exchange in 2003. The trading days with less than 500 transactions are discarded, because the return series is too short. We adopt the centred detrending moving average (DMA) approach to determine the Hurst indexes $H_{r}$, $H_{s}$, and $H_{x}$ of the time series of trade-by-trade returns, order signs, and relative prices of submitted orders for each trading day of each stock. A trading day is kept for further analysis if the estimation errors of the Hurst indexes $H_{r}$, $H_{s}$, and $H_{x}$ are all less than 0.02. If the number of trading days left for further analysis is less than 10, we exclude the stock from further analysis. As shown in Table~\ref{TB:Rho:Hr:Hs:Hx:Emp}, three B-share stocks are discarded. For a same company, the number $N_{\rm{day}}$ of trading days is larger for its A-share stock than its B-share stock (say, 000002 versus 200002). This is due to the fact that A-share stocks are usually much more actively traded than B-share stocks. The correlation coefficients $\rho(H_{r},H_{s})$ and $\rho(H_{r},H_{x})$ and their corresponding $p$-values are presented in Table~\ref{TB:Rho:Hr:Hs:Hx:Emp}.

\setlength\tabcolsep{2.5pt}
\begin{table}
  \centering 
  \caption{\label{TB:Rho:Hr:Hs:Hx:Emp} Empirical analysis of the cross-correlations between the Hurst index $H_r$ of the trade-by-trade returns and the Hurst indexes $H_s$ and $H_x$ of the order signs and the relative prices. Panel A and Panel B show respectively the results of A-share stocks and B-share stocks with the codes given in the first column. The second column presents the number of trading days $N_{\rm{day}}$ included in the analysis for each stock.}
  \medskip
\begin{tabular}{cccccc}\hline\hline
  Stock & $N_{\rm{day}}$ & $\rho(H_{r},H_{s})$ & $p$-value & $\rho(H_{r},H_{x})$ & $p$-value\\\hline
  \multicolumn{5}{l}{Panel A: A-share stocks}\\\hline
  000001&234&0.600&0.000&0.170&0.009\\
  000002&221&0.391&0.000&-0.268&0.000\\
  000009&218&0.643&0.000&-0.012&0.860\\
  000012&144&0.440&0.000&-0.209&0.012\\
  000016&146&0.487&0.000&-0.113&0.174\\
  000021&216&0.390&0.000&-0.017&0.806\\
  000024&99&0.423&0.000&0.130&0.200\\
  000027&164&0.476&0.000&0.026&0.739\\
  000063&167&0.422&0.000&-0.108&0.166\\
  000066&175&0.356&0.000&-0.048&0.525\\
  000088&56&-0.028&0.839&-0.207&0.125\\
  000089&131&0.320&0.000&-0.021&0.811\\
  000406&180&0.464&0.000&0.096&0.200\\
  000429&59&0.533&0.000&-0.105&0.427\\
  000488&87&0.153&0.157&0.084&0.439\\
  000539&66&0.313&0.011&0.065&0.607\\
  000541&24&-0.197&0.356&-0.489&0.015\\
  000550&176&0.363&0.000&0.052&0.493\\
  000581&60&0.363&0.004&0.212&0.105\\
  000625&192&0.257&0.000&-0.048&0.513\\
  000709&137&0.413&0.000&0.078&0.364\\
  000720&96&0.248&0.015&0.076&0.462\\
  000778&106&0.320&0.001&-0.067&0.497\\
  000800&231&0.332&0.000&-0.256&0.000\\
  000825&193&0.527&0.000&-0.031&0.665\\
  000839&237&0.277&0.000&-0.062&0.338\\
  000858&192&0.455&0.000&-0.036&0.625\\
  000898&205&0.477&0.000&0.066&0.346\\
  000917&119&0.404&0.000&0.032&0.733\\
  000932&188&0.504&0.000&0.069&0.348\\
  000956&227&0.373&0.000&-0.009&0.895\\
  000983&130&0.310&0.000&0.020&0.818\\\hline
  Aggregate&4876&0.326&0.000&-0.094&0.000\\\hline
\multicolumn{5}{l}{Panel B: B-share stocks}\\\hline
  200002&18&0.502&0.034&-0.326&0.187\\
  200012&20&0.184&0.439&-0.058&0.807\\
  200024&12&0.555&0.061&-0.016&0.961\\
  200429&15&0.689&0.004&0.066&0.815\\
  200488&95&0.339&0.001&0.200&0.052\\
  200539&76&0.517&0.000&0.084&0.470\\
  200550&51&0.413&0.003&-0.131&0.360\\
  200625&127&0.327&0.000&-0.139&0.120\\\hline
  Aggregate&414&0.397&0.000&-0.050&0.308\\\hline\hline
\end{tabular}
\end{table}

\begin{figure*}
  \centering
  \includegraphics[width=0.95\textwidth,height=0.72\textwidth]{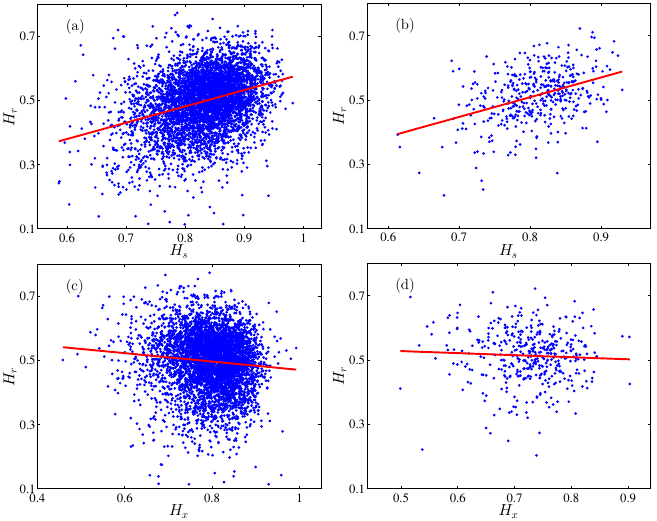}
  \caption{ Dependence of $H_r$ on $H_s$ and $H_x$ for A-share stocks and B-share stocks. (a) Dependence of $H_r$ against $H_s$ for the 32 A-share stocks. The Pearson correlation coefficient $\rho_{\rm{A}}(H_{r},H_{s})$ is 0.326 and the $p$-value is less than 0.1\%. (b) Dependence of $H_r$ against $H_s$ for the 8 B-share stocks. The Pearson correlation coefficient $\rho_{\rm{B}}(H_{r},H_{s})$ is 0.397 and the $p$-value is less than 0.1\%. (c) Dependence of $H_r$ against $H_x$ for the 32 A-share stocks. The Pearson correlation coefficient $\rho_{\rm{A}}(H_{r},H_{x})$ is -0.094 and the $p$-value is less than 0.1\%. (d) Dependence of $H_r$ against $H_x$ for the 8 B-share stocks. The Pearson correlation coefficient $\rho_{\rm{B}}(H_{r},H_{x})$ is -0.050 and the $p$-value is 0.308.}
  \label{Fig:Hr:Hs:Hx:Emp}
\end{figure*}

We first consider the correlation coefficients $\rho(H_r,H_s)$ between $H_r$ and $H_s$. For the A-shares, there are 30 stocks having positive correlation coefficients $\rho(H_r,H_s)$ and 2 stocks (000088 and 000541) with negative correlation coefficients. The two negative correlation coefficients are not significant even at the level of 10\%. The positive correlation between $H_r$ and $H_s$ of stock 000488 is also not significant. The positive correlation is significant at the level of 2\% for the remaining 29 stocks. For the B-shares, all the correlation coefficients are positive and 6 of them are significant at the 5\% level. If we aggregate all the data for the A-share stocks, we obtain an average correlation coefficient $\rho_{\rm{A}}(H_r,H_s)=0.326$ with the $p$-value less than 0.1\%, as also illustrated in Fig.~\ref{Fig:Hr:Hs:Hx:Emp}(a). Similarly, we obtain the average correlation coefficient $\rho_{\rm{B}}(H_r,H_s)=0.397$ for the B-share stocks, as illustrated in Fig.~\ref{Fig:Hr:Hs:Hx:Emp}(b). These observations indicate that $H_r$ is positively correlated with $H_s$, which verifies nicely the prediction of the computational experiments where $\rho(H_r,H_s)=0.953$.

We now consider the correlation coefficients $\rho(H_r,H_x)$ between $H_r$ and $H_x$. For the A-shares, there are 14 stocks having positive correlation coefficients and 18 stocks having negative correlation coefficients. There are only 5 correlation coefficients are significant at the 2\% level (4 negative and 1 positive), while the $p$-values of all other coefficients are greater than 10\%. When we aggregate all trading days, we obtain an average correlation coefficient $\rho_{\rm{A}}(H_r,H_x)=-0.094$ with the $p$-value less than 0.1\%, as illustrated in Fig.~\ref{Fig:Hr:Hs:Hx:Emp}(c). For the B-shares, there are 3 stocks with positive correlation coefficients and 5 stocks with negative correlation coefficients. All these correlations are not significant at the 5\% level. The average correlation coefficient is $\rho_{\rm{B}}(H_r,H_x)=-0.050$, which is also not significant. The scatter plot is illustrated in Fig.~\ref{Fig:Hr:Hs:Hx:Emp}(d). Overall, we find that there is a weak negative correlation between $H_r$ and $H_x$, which supports the prediction of the computational experiments where $\rho(H_r,H_x)=-0.156$.

We further perform linear regressions for the 32 A-share stocks and 8 B-share stocks using the following equation
\begin{equation}
  H_r = \beta_0+\beta_xH_x + \beta_s H_s.
  \label{Eq:Hr:Hx:Hs:Linear:Emp}
\end{equation}
For the A-share stocks, we obtain that $\beta_0=0.21$, $\beta_x=-0.33$ and $\beta_s= 0.66$. The adjusted $R^2$ is 0.156 and all the coefficients are significant at the 1\% level. For the B-shares, we obtain that $\beta_0=0.13$, $\beta_x=-0.10$ and $\beta_s= 0.57$ with the $p$-values being 0.052, 0.057 and 0.000, respectively. The adjusted $R^2$ is 0.140. These results are again in good agreement with Eq.~\ref{Eq:Hr:Hx:Hs:Linear} of the computational experiments. Note that we do not investigate the impact of the relative price distribution since there is no evidence showing that the relative prices are distributed as a Student function for the real LOB data \citep{Gu-Zhou-2009-EPL}.

\section{Discussion}

In this work, we have performed computational experiments to investigate the impacts of order flow characteristics on the correlation structure in return time series. The statistical properties of the relative prices of placed orders are found to have statistically significant but quantitatively negligible impacts on the Hurst index of returns. A fatter distribution of relative prices means more order are market orders with the prices larger than the opposite best prices. Since the order size is 1 in the MMF model, an order will not have persistent impacts on price movements. Hence, the parameter $\alpha_x$ only have negligible impacts on the persistence of returns quantified by $H_r$.

In contrast, the long memory effect in order direction plays a dominating role on the Hurst index of returns. The remarkable impact of the long memory feature of order direction can be explained. The Hurst index of the time series of order direction quantifies the degree of herding in order placement. Stronger herding incurs stronger persistence in the changes of prices. Moreover, the persistence of price changes is more evident because the order size is fixed to 1 and each order will be very likely to move the mid-price. These predictions from the computational experiments are in excellent agreement with empirical analysis on real order flow data.

Both computational experiments and empirical analysis show that the Hurst index of trade-by-trade return time series can deviate significantly from $H=0.5$. The intraday returns can be either antipersistent, persistent, or uncorrelated. This finding shows the evolutionary feature of traders' heterogenous behaviors. The deviation of $H_r$ from 0.5 does not necessarily imply that there are arbitrage opportunities at the micro scales and that the market is inefficient. Market friction such as transaction costs may eliminate the possibility to gain positive profits from such a predictability. In addition, the $T+1$ trading rule makes it more unlikely to make profits based on this finding of autocorrelations in the returns. The $T+1$ trading rule also provides a possible interpretation of the fact that the deviation is more severe in real stocks than in artificial stocks. The deviation of $H_r$ can also be understood in the sense that it takes time for the market to digest information.

Our findings imply that the correlation structure of return time series is not necessary to be stable over time. In particular, at small timescales, the Hurst index of returns can fluctuate following the fluctuations in the herding strength of order directions. We can conjecture that the return time series may exhibit long memory if the directions of placed orders are strongly correlated and become antipersistent if order directions are less correlated. These findings provide latent supportive evidence to the Adaptive Markets Hypothesis proposed by \cite{Lo-2004-JPM}.

We would like to note that the MMF model is very useful in the cost benefit analysis of market policies, as already done in \cite{Hayes-Paddrik-Todd-Yang-Beling-Scherer-2012-Conf}. For instance, there is a price limit rule in the Chinese stock market. The daily price limits are $\pm10\%$ for common stocks and $\pm5\%$ for specially treated stocks. A further modified MMF model shows that the price dynamics will be very different for asymmetric price limit setting. However, a detailed analysis of the topic is beyond the scope of this paper, which will be reported in a future work.

\begin{acknowledgements}
Zhi-Qiang Jiang, Gao-Feng Gu and Wei-Xing Zhou received support from the National Natural Science Foundation of China (71501072) and the Fundamental Research Funds for the Central Universities. Xiong Xiong and Wei Zhang received support from the National Natural Science Foundation of China (71532009,71131007) and the Program for Changjiang Scholars and Innovative Research Team in University (IRT1028). Wei Chen received support from the National Natural Science Foundation of China (71571121).
\end{acknowledgements}

\bibliographystyle{spbasic}      
\bibliography{E:/Papers/Auxiliary/Bibliography}

\end{document}